\UseRawInputEncoding
\documentclass[a4paper,10pt,oneside]{article}
\usepackage{icad2026,amsmath,epsfig,times,url,hyperref}
\usepackage{booktabs}
\usepackage{authblk}
\usepackage{siunitx}
\title{\textbf{SINGING MATERIALS: INITIAL EXPERIMENTS IN APPLYING SONIFICATION TO PHONON SPECTRA}}
\date{}

\author[1]{Lucy Whalley}
\author[2]{Rose Shepherd}
\author[3]{Jorge Boehringer}
\author[4]{Shelly Knotts}
\author[4]{Paul Vickers}
\author[2]{George Caselton}
\author[5]{Carol Davenport}
\author[2]{Christopher Harrison}
\author[3]{Bennett Hogg}
\author[5]{Antonio Portas}
\author[1]{Daniel Ratliff}
\affil[1]{School of Engineering, Physics and Mathematics, Northumbria University, UK}
\affil[2]{School of Mathematics, Statistics and Physics, Newcastle University, UK}
\affil[3]{School of Arts and Cultures, Newcastle University, UK}
\affil[4]{School of Computer Science, Northumbria University, UK}
\affil[5]{Faculty of Science and Environment, Northumbria University, UK}

\begin{document}
\ninept

\maketitle

\begin{sloppy}

\begin{abstract}

Solid materials may appear static, but at the atomic scale they are in constant vibrational motion. These vibrations, described by phonons, govern many key material properties, including structural stability, mechanical strength, optical behaviour, and thermal transport. Understanding phonon physics is therefore central to the rational design of materials with targeted functionalities. In this work, we introduce \texttt{SingingMaterials}, a modular Python package for sonifying materials simulation data. The software interfaces with the Materials Project database and is designed to be extensible, enabling the incorporation of additional sonification strategies and data sources. Built using the Sonification Toolkit \texttt{Strauss}, the current implementation supports three core approaches: spectral, synthesised, and sample-based. We demonstrate these approaches using phonon density-of-states data and evaluate their effectiveness through a user study, investigating whether listeners can distinguish differences in material properties from their auditory representations. The results show that sonification can provide an interpretable and complementary approach for exploring vibrational materials data.

\end{abstract}

\section{Introduction}
\label{sec:intro}

Materials that appear static at the macroscopic scale are, at the atomic level, in constant motion. 
The constituent atoms vibrate, and these vibrations play a fundamental role in determining a range of material properties, including structural stability, mechanical strength, optical response, and thermal transport \cite{Dove1993}. Understanding these atomic-scale motions is therefore essential for designing materials with specific properties, for example materials that absorb sunlight more efficiently in solar cells, or semiconductors that enable faster electronic switching \cite{Frost2017,Huang2025}.

In crystalline solid materials, where the atoms are arranged in a repeating pattern, atomic vibrations can be described using the quantum-mechanical concept of the phonon quasi-particle (Figure \ref{fig:phonons}).
Phonons represent collective lattice vibrations and can, in general terms, be thought of as `particles of heat'. 
This well-established framework allows the complex motion of many atoms to be understood in terms of a series of vibrational modes, each associated with a well-defined frequency \cite{Dove1993}.
The spectral nature of phonons makes them candidates for audification, in which frequency-domain information is mapped directly to audible sound \cite{Dombois2011}. Phonons also offer a conceptually intuitive basis for sonification more generally, as sound in solids is carried by lattice vibrations, and these vibrations are precisely what phonons describe.

The phonon spectrum of a material depends on both its atomic structure (the positions of atoms) and its chemistry (the chemical species). 
As a result, phonon spectra act as a vibrational fingerprint for materials. 
They can be measured experimentally using techniques such as Raman and infrared (IR) spectroscopy, and can also be predicted computationally using methods from quantum chemistry. 
Phonon spectra change with variations in temperature or pressure, and can vary spatially due to nanostructuring, interfaces, or atomic-scale defects. 
Experimental techniques such as micro-Raman spectroscopy can probe these spatial variations through changes in the Raman-active modes \cite{Ledinsky2015}.
Furthermore, advances in high-throughput computation have recently made it possible to calculate the phonon spectra for thousands of materials, and several open databases now provide access to this data \cite{Jain2013,Petretto2018}.
The scale, complexity, and wide availability of phonon data suggests that sonification could provide a useful complementary approach for the atomic-scale analysis of material behaviour.

This paper investigates how sonification can be applied to phonon spectra to create representations that are both informative and comfortable to listen to. 
We focus on the phonon density-of-states data (introduced in Section \ref{sec:dos}), as this is widely calculated and a large amount of openly available datasets already exists. 
The primary audience for these sonifications are researchers working in physics and materials science, with limited or no prior knowledge of sonification.
The sonifications are implemented in \texttt{SingingMaterials}, which is based on the open-source Sonification Toolkit \texttt{Strauss} \cite{SingingMaterials,Trayford2023,Trayford2025}.
\texttt{Strauss} has previously been applied to datasets in astrophysics; this is, to our knowledge, the first application beyond this field for use in a research context. The work presented here is intended as a pilot study demonstrating the potential of sonification in materials research, and as an invitation to the broader materials modelling community to explore these techniques.

\begin{figure}[tb]
\includegraphics[width=1.0\columnwidth]{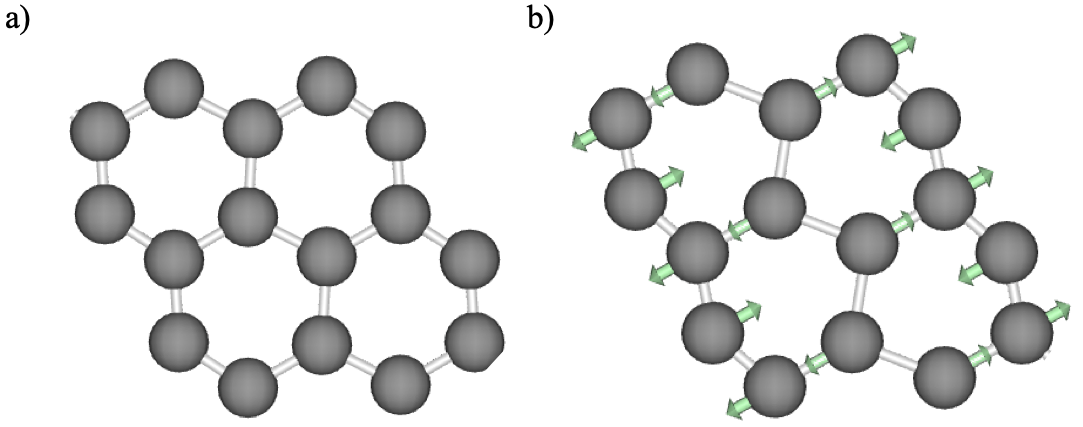}
\caption{Illustration of atomic vibrations in a crystalline lattice.
(a) Equilibrium (lowest energy) atomic positions in a 2D honeycomb structure.
(b) Example of a collective displacement pattern (phonon mode), where the green arrows indicate the displacement vectors of each atom. Each phonon mode is associated with a single vibrational frequency. In the harmonic approximation the full vibrational behaviour of a material arises from the superposition of many such modes. Figure generated using \url{https://henriquemiranda.github.io/phononwebsite}.}
\label{fig:phonons}
\end{figure}

\section{A Brief Survey of Sonification Applied to Materials Vibration Data}
\label{sec:litreview}

To assess how sonification has previously been applied to materials simulation data, a literature search was conducted using the OpenAlex scholarly database \cite{Priem2022}. 
Search terms related to atomistic modelling and vibrational spectroscopy (\textit{phonon}, \textit{molecular dynamics}, \textit{infrared}, \textit{Raman}) were combined with terms used in the auditory display community (\textit{sonification}, \textit{auditory display}). Only research articles were considered, and results were manually filtered to remove alternative uses of each word. In particular, the term \textit{phonon\textit} is sometimes used as the name of unrelated computational methods (e.g. \textit{phonon tracing}), while \textit{sonification} may appear in the context of sonication or ultrasound processing.

Searching for \textit{phonon} combined with \textit{sonification} or \textit{auditory display} yielded only two works. 
The first work approaches phonon sonification from an artistic and philosophical perspective. In this work, a method is proposed to use vibrational information derived from light-scattering measurements associated with an artwork to generate sound, with phonon-related data serving as the basis for musical composition \cite{TribotLaSpier2023}. The paper is not available in English, and only the abstract was inspected, however it appears that this method is speculative and not implemented or applied in practice.
The second relates to molecular dynamics simulations of polymer chains \cite{Lv2017Sonification}. 
In addition to conventional analysis, the authors employ sonification of the heat-flux signal in order to identify patterns in the phonon contributions to thermal transport.
Listening to these signals allows correlations in the heat flux to be perceived more easily than through visual inspection alone. Using this approach, the authors identify that low-frequency phonon modes play a dominant role in the observed thermal conductivity. This study demonstrates that sonification can serve as an exploratory analysis tool for complex materials simulation data, helping researchers detect temporal structures and correlations in atomic-scale dynamics that may otherwise be difficult to identify. To our knowledge, this appears to be the only work applying sonification to atomistic simulations of phonon-related transport processes. 

Expanding the search to \textit{molecular dynamics}, a closely related simulation method that provides insight into vibrational behaviour, returned twelve additional articles involving \textit{sonification} or \textit{auditory display}. Of these, three relate to crystalline materials. 
The first is the Aural Structures project, which explores the use of sonification to create sonic representations of crystal structures for educational purposes \cite{Eramo2018AuralStructures}. 
Physical, chemical, and structural properties of atoms within a crystal are mapped to sound parameters such as pitch, duration, timbre, and dynamics. 
The resulting sounds aim to help learners understand concepts such as polymorphism, solid solutions and molecular disorder.
The authors also describe the development of software capable of automatically generating a musical score from crystallographic data (e.g. chemical composition and space group). 
While the work focuses on structural properties rather than vibrational dynamics (phonons), it represents an example of how atomistic structural data from crystalline materials can be translated into sound for educational purposes.
The second paper describes an interactive sonification of molecular dynamics simulations, aimed at creating what the authors describe as `molecular musical instruments' \cite{Mitchell2020MolecularMusicalInstruments}. 
The method converts real-time simulation data into sound using scanned synthesis, a sound synthesis technique that maps the geometric motion of a system onto audio waveforms. 
It is applied to both molecular systems and low-dimensional crystalline materials (graphene and carbon nanotubes). The approach is implemented in a standalone application and audio plug-ins, and focuses on interactive and artistic exploration of molecular simulations, rather than the analysis of vibrational spectra or phonon properties.
The third paper is a commentary piece introducing two articles based on the idea of hierarchical design structures \cite{Guo2019NaturesWay}. It references in passing an article which explores the combined use of artificial intelligence and sonification to explore the design principles of hierarchical structures across domains from molecular materials to music \cite{Yu2019SelfConsistentSonification}. This paper is described below.

The remaining nine studies focus on molecular systems only. 
Molecules lack the translational symmetry required for the existence of phonons, however their dynamics can be studied at the atomic scale using similar simulation techniques. 
As such, two papers particularly relevant to this project will be described. 
The first paper introduces a self-consistent sonification method for protein vibrations \cite{Yu2019SelfConsistentSonification}. Vibrational frequencies are transposed into the audible range, alongside other sonic parameters (duration, amplitude) derived from other properties of the vibrational modes. This creates musical scores that preserve the physical relationships present in the protein. A neural network is trained on these scores and used to generate musical compositions, which are then in turn translated into new protein sequences.  
The work illustrates how sonification might be used to encode vibrational information, and highlights possible applications in both scientific exploration and artistic composition.
The second work outlines an approach for sonifying the IR vibrational spectra of small molecules \cite{Kim2026MusicalMolecules}. 
This maps experimentally measured vibrational frequencies into the audible range using an anharmonic oscillator model. By comparing harmonic and anharmonic mappings, the authors demonstrate how features of molecular vibrational spectra, such as frequency shifts, beating patterns, and combination bands, produce distinctive audible characteristics.
The work also introduces a time-dependent model of intramolecular vibrational energy redistribution. In this model, an individual vibrational mode is “plucked”, allowing energy to redistribute among other modes over time, producing evolving sonic textures that reflect the dynamics of energy flow within the molecule. The authors suggest that sonification of IR spectra could provide an intuitive and pedagogical representation of vibrational phenomena.

This brief survey suggests that sonification has been explored in a number of contexts related to atomic-scale structure and dynamics.
In most cases simulation or spectroscopy data is used to generate sound for educational purposes or artistic exploration (e.g. \cite{TribotLaSpier2023,Eramo2018AuralStructures,Mitchell2020MolecularMusicalInstruments,Kim2026MusicalMolecules}), with only two articles identified which use sonification as a research tool for understanding existing materials or discovering new materials \cite{Lv2017Sonification,Yu2019SelfConsistentSonification}. 
In particular, the phonon quasi-particle, which provides the framework for describing vibrations in crystalline materials, has had very limited use as the basis for sonification \cite{Lv2017Sonification}.
This gap is notable given the conceptual connection between phonons and sound, and the increasing availability of large-scale phonon datasets produced by modern computational materials science.

\section{Phonon density-of-states data}
\label{sec:dos}

\begin{figure}[tb]
\includegraphics[width=1.0\columnwidth]{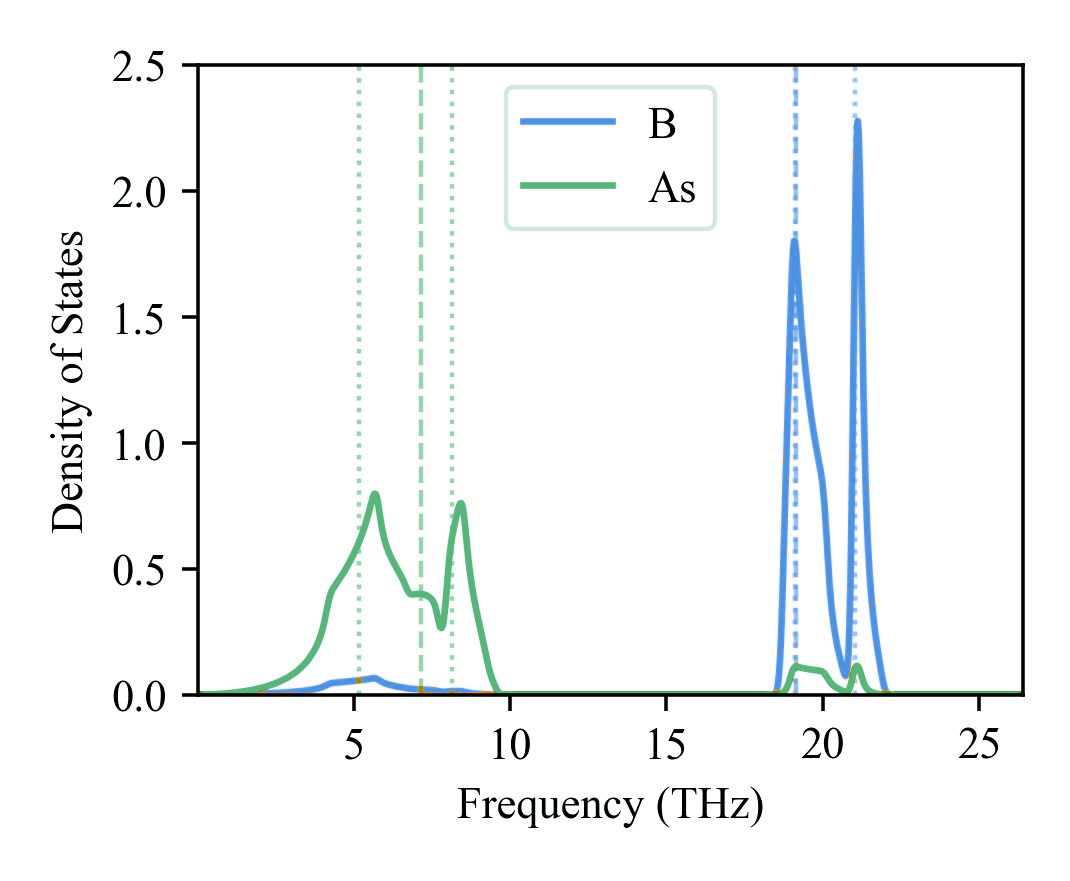}
\caption{Phonon projected density-of-states (PDOS) for boron arsenide (BAs). The x-axis shows the phonon frequency (in THz), and the y-axis shows the PDOS (in arbitrary units). The species-resolved phonon band centres (dashed lines) reflect a separation between lower-frequency As vibrations and high-frequency B vibrations. The interquartile ranges (dotted lines) show that As exhibits a broader spread of phonon frequencies.}
\label{fig:BAs_dos}
\end{figure}

In this work we explore the sonification of phonon density-of-states (DOS) data. 
The phonon DOS provides a summary of the vibrational behaviour of a crystalline material, describing how many vibrational states occur within a given frequency range. 
Peaks in the DOS therefore indicate frequency ranges where many vibrational modes occur. 
In some materials, frequency regions with no allowed vibrational states may also appear; these are known as phonon band gaps.
The phonon DOS is widely calculated because it determines a range of thermodynamic properties including free energy, entropy, and heat capacity, and therefore plays an important role in assessing material stability. It can also provide useful insight into thermal transport.

In addition to the total phonon DOS, it is often useful to consider a projected-DOS (PDOS). 
Here, the vibrational contributions of each atomic species are separated, which allows their contribution to the phonon spectra to be distinguished. 
Figure \ref{fig:BAs_dos} illustrates an example phonon PDOS for the crystalline material boron arsenide (BAs).
BAs is a material with exceptional thermal, electronic, and optical properties, which makes it a highly promising material for next-generation electronics and quantum computing technologies. Its exceptional thermal transport properties are associated with the significant phonon band gap shown in Figure \ref{fig:BAs_dos}.
We emphasise that the shape of the phonon DOS or PDOS varies widely between materials: some spectra extend over relatively narrow frequency ranges (e.g. up to $\sim$2 THz), others exhibit broad features rather than distinct peaks, and some show no phonon band gap.

The large mass difference between boron (B) and arsenic (As) leads to two distinct regions: a lower-frequency region dominated by As vibrations and a higher-frequency region dominated by B vibrations. This behaviour can be understood from the highly simplified model of a harmonic oscillator: 
\begin{equation}
   \omega = \sqrt{\frac{k}{m}} \label{spring} 
\end{equation}
where $\omega$ is the phonon frequency,  $k$ represents the bond stiffness, and $m$ is the atomic mass. For comparable bond strengths, lighter atoms therefore tend to vibrate at higher frequencies, while heavier atoms tend to dominate the lower-frequency region of the spectrum.

\begin{figure*}[tb]
\centering
\includegraphics[width=0.8\textwidth]{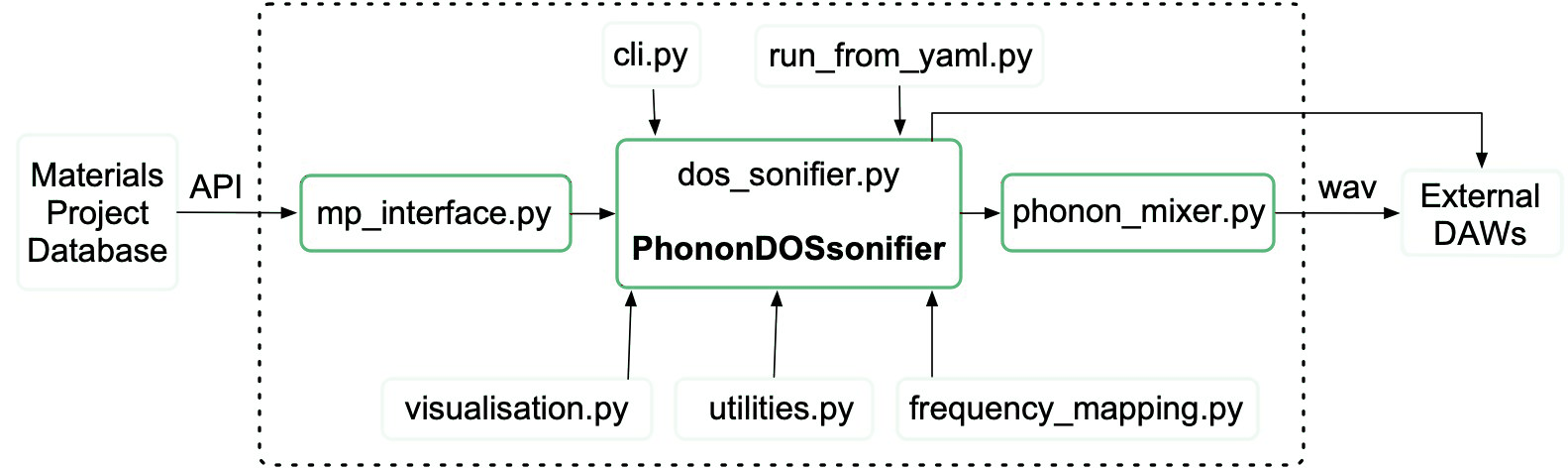}
\caption{Architecture of the \texttt{SingingMaterials} package. Phonon data is retrieved from the Materials Project Database via its API and processed within a modular Python workflow. The central object is the \texttt{PhononDOSSonifier} class which implements sonification methods using the \texttt{Strauss} toolkit. Sonifications are output as \texttt{.wav} files.}
\label{fig:technology}
\end{figure*}

In order to extract summary quantities from the phonon DOS, several statistical descriptors can be defined.
One widely used measure is the phonon band centre, which is the average phonon frequency ($\omega$) weighted by the DOS ($g(\omega)$):
\begin{equation}
\bar{\omega} = \frac{\int_0^\infty \omega \, g(\omega)\, d\omega}{\int_0^\infty g(\omega)\, d\omega}. \label{pbc}
\end{equation}
A temperature-dependent phonon band centre can also be defined by weighting with both the DOS and the Bose-Einstein occupation factor $n(\omega,T)$:
\begin{equation}
\bar{\omega}(T) = \frac{\int_0^\infty \omega \, g(\omega)\, n(\omega,T)\, d\omega}{\int_0^\infty g(\omega)\, n(\omega,T)\, d\omega}. \label{wpbc}
\end{equation}
Here, $n(\omega,T)$ accounts for the thermal population of vibrational modes, such that the relative contribution of higher-frequency modes increases with increasing temperature.
However, it should be noted that other temperature-dependent phenomena, such as thermal expansion of the lattice, are not captured in this simplified description.

A second statistical measure is the interquartile range (IQR) of the frequency distribution. 
The IQR provides a measure of the spread of vibrational frequencies within the spectrum, indicating whether the vibrational states are broadly distributed or concentrated within a narrower frequency range. This information can help characterise the degree of localisation in the vibrational spectrum. 
For example, an atomic-scale defect may produce a localised `rattling' vibration which appears in the phonon DOS as a narrow peak \cite{Stoneham1975}.

\section{Technological approach}
\label{sec:approach}

Figure \ref{fig:technology} illustrates the architecture of the open source \texttt{SingingMaterials} Python package and the interaction between its components \cite{SingingMaterials}. The core Python modules are listed below, followed by a discussion of software design decisions. 

\begin{itemize}
    \item \texttt{mp\_interface} retrieves and post-processes phonon DOS data from the Materials Project database.
    \item \texttt{dos\_sonifier} implements the \texttt{PhononDOSSonifier} class along with associated sonification methods, outputing the sonifications as \texttt{.wav} files
    \item \texttt{phonon\_mixer} optionally concatenates and overlays the \texttt{.wav} files to produce a final sonification
\end{itemize} 

\subsection{Python-based workflows}
\label{ssec:python}

The target users of this work are computational materials researchers, and Python is the dominant programming language in this community. 
Recent research has similarly highlighted Python as a common platform for sonification workflows applied in a scientific context, with several general-purpose sonification packages implemented in this language \cite{VallejoBudziszewski2025}. 
By adopting Python, this project aligns with established practices in materials modelling and reduces the barrier to adoption.
It also allows \texttt{SingingMaterials} to more easily integrate with existing Python libraries that underpin this work, specifically \texttt{Strauss} and the Materials Project API.

Beyond these domain specific Python libraries, \texttt{SingingMaterials} relies only on the standard scientific Python stack (\texttt{NumPy}, \texttt{Pandas}, \texttt{SciPy}, \texttt{Matplotlib}) and the \texttt{ffmpeg} command-line tool. The latter is used for combining audio outputs, providing functionality not currently available within version one of \texttt{Strauss}. This minimal Python-based dependency stack improves portability, ease of installation, and long-term maintainability.

\begin{figure}[tb]
\includegraphics[width=1.0\columnwidth]{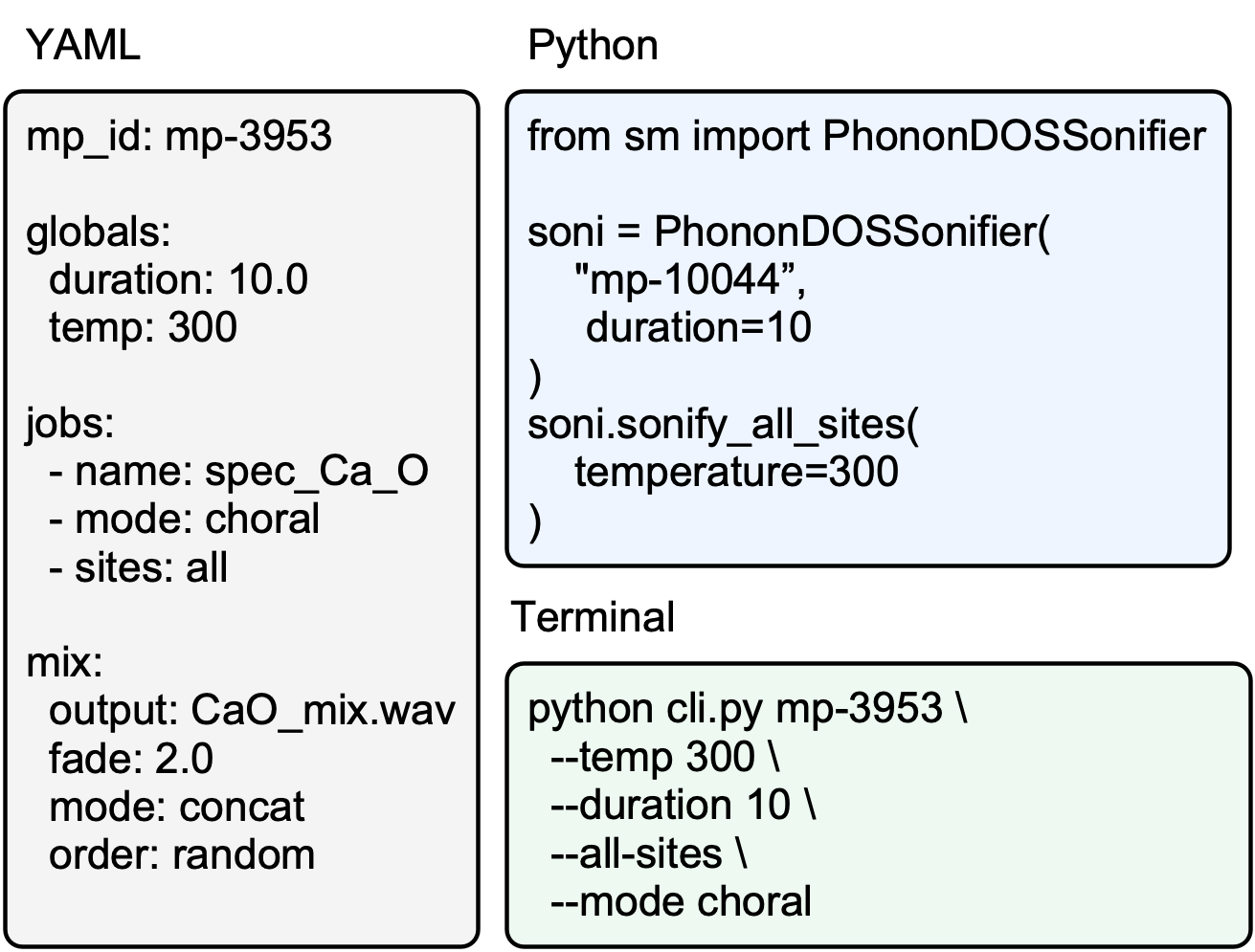}
\caption{Example usage of the three user interfaces for \texttt{SingingMaterials}. a) YAML file with sonification and mixing parameters specified; b) Python API for integration into existing workflows and the Jupyter ecosystem; c) command line interface for integration into remote and/or high-throughput workflows.}
\label{fig:ui}
\end{figure}

\subsection{Domain-specific interfaces}
\label{ssec:domain}

\texttt{SingingMaterials} is designed to interface directly with the Materials Project database, providing immediate access to the properties of over 200,000 materials and thus enabling researchers to apply sonification to their systems of interest \cite{Jain2013,Petretto2018}. This is achieved via the Materials Project API, which supports queries using unique material IDs and returns up-to-date data in a consistent format. To further enhance compatibility with established materials modelling workflows, future development will include integration with \texttt{Phonopy}, a widely used Python-based tool for phonon calculations \cite{Togo2015}.

\subsection{User interfaces}
\label{ssec:interface}

\texttt{SingingMaterials} provides multiple modes of interaction to support different use cases. 
Computational materials scientists are typically familiar with script-based workflows; 
the design of \texttt{SingingMaterials} therefore emphasises accessibility through a programmatic API, enabling users to integrate sonification into existing pipelines. 
A command-line interface provides a simplified entry point for common tasks. 
In addition, workflows can be fully specified using a YAML configuration file, which defines parameters such as the Materials Project ID, temperature, sonification method, and mixing settings. This approach supports reproducibility and allows complete workflows to be easily shared. 
Code and input file examples are shown in Figure \ref{fig:ui}.
Future work will explore the development of a web-based interface to further broaden accessibility.

\subsection{Sustainable software design}
\label{ssec:sustainable}

\begin{table*}[tb]
\centering
\caption{Parameter mappings used to display the phonon density-of-states (DOS) as audio. The final column specifies the sonification approaches in which each mapping was applied.}
\label{table:mappings}
\begin{tabular}{llll}
\toprule
\textbf{Phonon DOS quantity} & \textbf{Audio parameter} & \textbf{Mapping direction}  &  \textbf{Approaches} \\
\midrule
 Band centre                  & Frequency        & Higher value $\rightarrow$ higher pitch & Synthesised, sample-based \\
 Density of states            & Amplitude    & Larger value $\rightarrow$ louder sound & Synthesised, sample-based \\
 Interquartile range (IQR)    & Modulation   & Broader spread $\rightarrow$ more fluctuation (tremolo/vibrato) & Synthesised  \\
\bottomrule
\end{tabular}
\end{table*}

Computational materials scientists are active contributors to community-developed codes including those specifically developed for post-processing and visualising materials modelling data (for example, \texttt{sumo} \cite{Ganose2018}).
\texttt{SingingMaterials} is designed as a modular and extensible package, allowing new sonification methods and data interfaces to be incorporated with minimal modification of the existing code base. 
The project is developed openly on GitHub, encouraging community contributions and reuse. 
Tutorials and example workflows are provided through Jupyter notebooks hosted in the project repository.
Ongoing development priorities include expanding test coverage, improving API documentation, and providing additional tutorial workflows to support new users.

\section{Sonification strategies}
\label{sec:strategies}

\subsection{General approach}
\label{ssec:general}

Three complementary sonification strategies---spectral, synthesised, and sample-based---are implemented in \texttt{SingingMaterials}, each encoding different aspects of the phonon DOS data. 
All approaches are based on methods available within the \texttt{Strauss} sonification toolkit \cite{Trayford2023,Trayford2025}.
The synthesis and sample-based strategies use parameter mapping sonification, in which features of the phonon DOS are mapped onto auditory parameters. 
These produce a `chordal fingerprint' for each material,
with the phonon band centre of each atomic species corresponding to a characteristic frequency.
The third spectralisation approach uses an inverse Fast Fourier Transform (iFFT) to generate spectral audifications from data.
A listener study has previously demonstrated that this can be used to identify spectral features such as noise, line width, and flux ratios in galaxy spectra \cite{Trayford2023RASTI}. 
This work also highlighted the importance of listener calibration as participants were strongly influenced by the previous sound they had heard. 
The present study mitigates this effect by structuring each task as a paired comparison between two materials.

All approaches require mapping phonon frequencies, typically spanning multiple orders of magnitude (e.g. 0.1 to 10 THz), into a smaller audible range (e.g. 100 to 1500 Hz). Both linear-to-linear and logarithmic-to-logarithmic mappings are supported; the former preserves absolute frequency spacing, so that equal differences in vibrational frequency map to equal differences in audio frequency, whilst the latter preserves relative frequency relationships, so that equal ratios of vibrational frequency map to equal ratios of audio frequency.
Temperature effects can be incorporated by weighting the phonon DOS with the Bose-Einstein occupation factor (as in Equation \ref{wpbc}).
A mixing stage allows the outputs of different sonification strategies to be combined. 
This enables both comparative listening (e.g. between materials) and the construction of longer, more aesthetically engaging audio sequences. 
Standard audio processing techniques such as fade-in and fade-out are applied to improve listener comfort. Example sonifications are available online \cite{ExampleData}. 

\subsection{Spectral}
\label{ssec:spectral}
In the spectral approach, the phonon DOS is interpreted directly as a frequency-domain signal. 
The DOS is mapped onto a set of amplitudes across frequencies and transformed into a time-domain audio signal using an iFFT implemented in the \texttt{Strauss} generator \texttt{Spectraliser}.
This method preserves the overall spectral structure of the vibrational data, providing a direct representation of the phonon frequency distribution. Peaks in the DOS correspond to dominant frequency components in the resulting audio signal. 

\subsection{Synthesised}
\label{ssec:synth}
The synthesised mapping approach uses parameter mapping sonification to generate sound using the \texttt{Strauss} generator \texttt{synthesiser}. The phonon band centre of each PDOS (corresponding to an atomic species) is mapped to a distinct frequency, producing a chord-like auditory representation of the material. The IQR can be mapped to either a tremolo or vibrato low-frequency modulation. The mappings are summarised in Table \ref{table:mappings}.

\subsection{Sample}
\label{ssec:sample}
The sample-based approach also produces a chordal representation, but maps phonon frequencies onto a discrete musical scale using the \texttt{Strauss} generator \texttt{sampler}. The frequencies are binned and mapped to the nearest notes in a chromatic scale, which are used to trigger audio samples. 
While this sample-based approach reduces spectral resolution compared to the synthesised approach, subtle variations inherent to sampled sounds introduce a time-varying character that enhances the aesthetic quality of the sonification. 
This is particularly important for this work, as the underlying parameter mappings themselves produce largely static sounds that do not evolve in time.
The mappings are summarised in Table \ref{table:mappings}.

\section{Case studies}
\label{sec:studies}

To demonstrate usage of \texttt{SingingMaterials}, two case studies applied to a set of representative materials are considered. Each case study is a simplified version of a routine task in phonon analysis: tracking how mode frequencies evolve under changing conditions (the stiffness test) and identifying additional modes introduced by compositional changes or changes to symmetry (the mass test). 
The materials were selected to span a range of vibrational and physical properties. In addition, all selected systems are technologically relevant and actively studied within the materials science community. A summary of the materials is given in Table \ref{table:materials}, including their key physical and phonon properties. 

All three sonification approaches applied to each material are available online \cite{ExampleData}. The linear-to-linear mapping is used throughout; a logarithmic mapping may offer perceptual advantages, and a systematic comparison of mapping strategies is left to future work. Phonon frequencies are scaled into an audible range of 190 to 1500 Hz to promote listener comfort. The sample-based approach uses choral recordings freely available via the Pianobook community sample library \cite{SolitaryChoir}.


\begin{table*}[t]
\centering
\caption{Summary of materials used in the case studies. Properties were obtained from the Materials Project (MP), and phonon band centres were calculated using the \texttt{SingingMaterials} package. The mass ratio is defined as the ratio between the largest and smallest atomic masses in the material, and is not applicable for single-species materials. Gap refers to a frequency region in the density-of-states where no vibrational modes are present. Band centres are reported for each species in the order they appear in the chemical formula.}
\label{table:materials}
\begin{tabular}{l l c c c c ccc l}
\toprule
\textbf{Formula} & \textbf{MP ID} & \textbf{Mass ratio} & \textbf{Bulk modulus (GPa)} & \textbf{Gap} & \textbf{Max freq (THz)} & \multicolumn{3}{c}{\textbf{Band centres (THz)}} \\
\cmidrule(lr){7-9}
 &  &  &  &  &  & \textbf{1} & \textbf{2} & \textbf{3} &  \\
\midrule
SrTiO$_3$ & 4651  & 5.5 & 168 & yes & 25  & 5.8 & 10.9 & 13.5  \\
PbTe      & 19717 & 1.6 & 38  & no  & 3.7 & 1.75 & 2.6 & n/a  \\
C (diamond)   & 66    & n/a & 435 & no  & 40  & 29 & n/a & n/a  \\
Si        & 149   & n/a & 89  & no  & 15  & 9.9 & n/a & n/a  \\
BAs       & 10044 & 6.0 & 132 & yes & 22  & 19.1 & 7.1 & n/a  \\
MgO       & 1265  & 1.5 & 151 & no  & 20  & 10.4 & 13.1 & n/a  \\
\bottomrule
\end{tabular}
\end{table*}

\subsection{Material stiffness}
\label{ssec:bulk}

The first case study considers the relationship between vibrational properties and the bulk modulus, a measure of a material’s stiffness against uniform compression. In the model of a harmonic spring, vibrational frequency is related to bond stiffness through Equation \ref{spring}. Within this approximation, stiffer bonds correspond to higher vibrational frequencies. In real materials the bonding environment is anharmonic and more complex, however this simplified picture can often be used to rationalise the mechanical properties of a material.
Table \ref{table:materials} shows that both the maximum phonon frequency and the maximum phonon band centre correlate with the bulk modulus for the materials in this study. 
Representative examples include diamond (one of the many carbon allotropes), which exhibits both a high bulk modulus and high-frequency vibrational modes, and lead telluride (PbTe), which is comparatively soft and dominated by lower-frequency vibrations. Silicon (Si) provides an instructive intermediate case: although it shares the same crystal structure as diamond, its weaker bonding results in lower characteristic frequencies and a reduced bulk modulus.
Following the mappings in Table \ref{table:mappings}, we expect the audio representations of stiffer materials to exhibit a shift towards higher pitch.

\subsection{Mass difference}
\label{ssec:mass}

The second case study explores the effect of the mass difference between atomic species on the phonon DOS. Insight can be gained from the textbook example of a one-dimensional diatomic chain, in which alternating light and heavy atoms give rise to two distinct phonon branches \cite{Dove1993}. As the mass ratio increases, the separation between these branches widens, leading to the formation of a phonon band gap.
This behaviour motivates the use of the separation between species-resolved phonon band centres as a proxy for the mass difference in real materials. In practice, this is again a simplification, as bonding character and the exact structural configuration also influence the phonon spectrum.
Table~\ref{table:materials} shows that the presence or absence of a phonon gap is strongly associated with the  mass ratio across this set of materials.
Boron arsenide (BAs) provides a clear example, where the large mass ratio between B and As leads to a pronounced separation in the PDOS (Figure \ref{fig:BAs_dos}). In contrast, magnesium oxide (MgO), with a smaller mass ratio, exhibits no separation. A more complex case is the ternary material strontium titanate (SrTiO$_3$); here, both mass differences and variations in bonding character contribute to the observed vibrational structure.
Following the mappings in Table \ref{table:mappings}, we expect the audio representations of materials with a larger mass difference to have a larger separation in frequencies.

\section{User evaluation}
\label{sec:evaluation}

\begin{table}[t]
\centering
\caption{Accuracy of user responses (n = 26). Values represent the proportion of correct responses. p-values are computed using a two-sided binomial test against random guessing (p = 0.5). Different material systems were used for the stiffness and mass difference tasks. Statistically significant results (p $<$ 0.05) are shown in bold.}
\label{table:accuracy}
\begin{tabular}{lccccc}
\toprule
 & \multicolumn{2}{c}{\textbf{Stiffness}} & \multicolumn{2}{c}{\textbf{Mass difference}} \\
\cmidrule(lr){2-3} \cmidrule(lr){4-5}
\textbf{Category} & \textbf{Accuracy} & \textbf{p-value} & \textbf{Accuracy} & \textbf{p-value} \\
\midrule
\multicolumn{5}{l}{\textit{By method}} \\
Synth          & \textbf{0.795} & 0.002 & 0.679 & 0.076 \\
Sample-based   & \textbf{0.782} & 0.009 & 0.577 & 0.557 \\
Spectral       & \textbf{0.821} & 0.002 & \textbf{0.808} & 0.002 \\
\midrule
\multicolumn{5}{l}{\textit{By material system}} \\
Diamond--PbTe        & \textbf{0.821} & 0.002 & -- & -- \\
BAs--SrTiO$_3$       & \textbf{0.833} & 0.001 & -- & -- \\
Si--SrTiO$_3$        & \textbf{0.744} & 0.029 & -- & -- \\
Diamond--SrTiO$_3$         & -- & -- & \textbf{0.782} & 0.009 \\
BAs--MgO       & -- & -- & 0.667 & 0.169 \\
MgO--SrTiO$_3$       & -- & -- & 0.615 & 0.327 \\
\bottomrule
\end{tabular}
\end{table}

Participants were asked to rate their familiarity with the concepts of phonons and sonification on a 1--5 scale, with 1 described as \textit{`not at all familiar'} and 5 as \textit{`extremely familiar'}. 
Familiarity with phonons received a median score of 3.5 (IQR, 2.8-5.0), and familiarity with sonification 2.5 (1.0-3.3), indicating that the study group was representative of the intended audience.
The age distribution indicates that participants are early- to mid-career researchers, with the majority of participants in the 25--29 age range (12 participants), followed by 35--39 (6 participants) and 30--34 (4 participants), with smaller numbers in other age groups.
Participants completed the study in typical working environments, reporting some low-level background noise in four cases. Listening was primarily conducted using standard consumer devices (e.g. laptop speakers or headphones), reflecting realistic usage conditions for this audience.

The evaluation consisted of three parts. In the first, participants were presented with pairs of sonifications and asked to identify which corresponded to the stiffer material. This comparative format was chosen because judgments are difficult without a reference point \cite{Trayford2023RASTI}. Participants were not given explicit information about the mapping, but were informed that \textit{`increased material stiffness is associated with higher frequency phonons'}, reflecting a condition in which researchers interact with sonifications without detailed prior knowledge, for example if producing the sonification through a web application.

Performance on the stiffness task was consistently high across all sonification methods (Table \ref{table:accuracy}). All methods achieved accuracy significantly above random guessing (p $<$ 0.05), indicating that participants were reliably able to identify the stiffer material from the sonifications.
This provides evidence that the mapping from phonon frequency to pitch is intuitive, even without explicit explanation of the sonification design. The consistency of results across both methods and material systems suggests that this relationship is robust and interpretable by the target audience.

In the second task, participants were asked to identify which material exhibited the larger mass difference, based on paired sonifications. They were informed only that \textit{`materials with larger mass differences exhibit greater separation in their phonon frequencies'}. 
Performance on the mass difference task was more variable. Only the spectral method achieved accuracy significantly above random, while the synthesised approach showed borderline performance, and the sample-based approach did not perform significantly above chance.
At the level of individual material systems, only the comparison with the starkest mass difference (diamond-SrTiO$_3$) reached statistical significance, while others did not.

These results highlight a clear distinction between the two tasks. The stiffness task relies on a more direct and intuitive mapping between frequency and pitch, which listeners are able to use effectively.
In contrast, the mass difference task depends on a mapping which is not so immediately obvious.  
This is supported by a participant's free-text response: \textit{`It\textquotesingle s ambiguous how the difference in phonon frequency translates to the sound that I heard. I just assumed that it is the difference between the voices I hear at once.'} 
The mass-difference task also makes a harder perceptual demand as participants must compare the spacing between pitches for one material with the spacing for another, rather than detecting the direction of a pitch shift.
We note that with 26 participants statistical power was limited, and small or moderate effects may have gone undetected; the absence of a significant difference in the mass difference task should therefore not be interpreted as evidence that no effect exists.

In the third task participants were asked to rate listening comfort on a 1--5 scale with 1 described as \textit{`very uncomfortable'}, 3 as \textit{`neither uncomfortable nor comfortable'} and 5 as \textit{`very comfortable'}. Across all sonifications, the median comfort score was 3.0 (IQR, 2.0--4.0), indicating that the sounds were generally acceptable to listen to. 
The highest comfort ratings were observed for the sample-based method, 4.0 (3.0--5.0), followed by the synthesised method, 4.0 (3.0--4.3), while the spectral method was rated substantially lower, 2.5 (2.0--3.3). 
This data is supported by free-text responses: one participant wrote of the sample-based sonification, `\textit{Particularly enjoyed the choral sounds, these were quite meditative},' while another commented on the spectral method, `\textit{The whistling tones were quite harsh when high pitched or clashing notes.}' This suggests that timbral character influenced participants' comfort judgements.

When considered alongside the mass difference task, these findings reveal a trade-off between task accuracy and listening comfort. The spectral method, which most directly preserves the structure of the phonon data, achieved the highest performance for interpreting both stiffness and mass difference, but was generally less comfortable to listen to. 
In contrast, the synthesised and sample-based approaches, which introduce additional musicality through parameter mapping, were judged as more comfortable to listen to, but were less effective at conveying the mass difference relationship.
Overall, our findings reinforce the well-established idea that the choice of sonification strategy should be guided by the intended application, with a careful balance between analytical accuracy and user engagement \cite{Kramer1999}.

\section{Conclusions}

In this work, we have presented an initial investigation into how established sonification strategies can be applied to phonon data describing the atomic-scale vibrations of crystalline materials. To support this, we developed the \texttt{SingingMaterials} Python package, which is built on the \texttt{Strauss} sonification toolkit, to provide a flexible and extensible framework for exploring these approaches.
Results from a user evaluation study demonstrate that listeners are able to distinguish differences in material properties from sonified phonon data. In particular, accuracy for the stiffness task was significantly above random for all methods, whereas performance for the mass difference task was more variable, with only the spectral approach showing a significant result. 

These results provide a foundation for the further use of sonification in materials modelling. They highlight the importance of carefully balancing data fidelity and user engagement, and the importance of providing appropriate user guidance when conveying more complex physical relationships. Future work will explore additional sonification strategies for application to phonon data, and apply these approaches to time-dependent spectra in non-equilibrium systems. It would also be interesting to consider defect systems, as these are of significant interest in the materials science community, and introduce localised vibrational modes that are challenging to identify through standard visualisation techniques. 

\label{sec:conclusions}

\section{ACKNOWLEDGMENTS}
\label{sec:ack}

L.W. thanks Dr Robert Lieck at Durham University for discussion. L.W., J.B., S.K., P.V., C.D., C.H., B.H., A.P and D.R. gratefully acknowledge funding from UK Research and Innovation (UKRI) under the \textit{Sonic Intangibles: Northumbria and Newcastle Universities Sonification Hub for Innovation in Sound and Meaning} project, supported through the Cross Research Council Responsive Mode scheme (grant number MR/Z506448/1). R.S. is funded through STFC grant (NUdata Centre for Doctoral Training), ST/W006790/1. G.C. is funded through grant number MR/V022830/1.
\bibliographystyle{IEEEtran}
\bibliography{refs2026}
%
%
%
%

\end{sloppy}

\end{document}